\documentclass[a4paper]{jpconf}
\usepackage{graphicx}
 \usepackage{amssymb}
 \usepackage{amstext}

\begin{document}
\title{Testing Elementary Cycles Formulation of Quantum Mechanics in Carbon Nanotubes and Superconductivity}

\author{Donatello Dolce \& Andrea Perali}

\address{University of Camerino, Piazza Cavour 19F, 62032 Camerino, Italy.}


\begin{abstract}
Elementary Cycles are intrinsic periodic phenomena, classical in the essence, whose classical relativistic dynamics reproduce the complete coherence (perfect recurrences) typically associated to the pure quantum behaviours of elementary particles. They can be regarded as effective representations of 't Hooft Cellular Automata.  By means of Elementary Cycles physics we obtain a consistent, intuitive, novel derivation of the peculiar quantum dynamics of electrons in Carbon Nanotubes, as well as of Superconductivity fundamental phenomenology. In particular we derive, from classical arguments, the essential electronic properties of graphene systems, such as energy bands and density of states. Similarly, in the second part of the paper, we derive the Superconductivity  fundamental phenomenology in terms of simple geometrical considerations, directly from the Elementary Cycles dynamics rather than from empirical aspects and effective quantities connected to the microscopical characteristics of materials as in the standard approaches to Superconductivity. With this approach simple geometrical considerations about the competition between the quantum recurrence and the thermal noise allow for a novel interpretation of the occurrence of high temperature superconductivity and the related gauge symmetry breaking mechanism. 
\end{abstract}

\section{Introduction}

As proven by G. 't Hoof in recent works \cite{'tHooft:2001ar,Hooft:2014kka}, particular classical systems, such as Cellular Automata (CA), have much in common with Quantum Mechanics (QM). This correspondence has been largely tested and extended by the Elementary Cycles (ECs) theory \cite{Dolce:SuperC,Dolce:EPJP,Dolce:cycles,Dolce:2009ce,Dolce:tune,Dolce:AdSCFT,Dolce:EC2CA,Dolce:ECunif} which has originated a novel consistent formulation QM in all the its fundamental aspects. For the scope of this paper an EC can be represented as a (massless) particle moving very fast on a circle with time period $T$. In order to represent the electrons in graphene systems, such an EC can also be assumed on a lattice, similarly to a periodic 't Hooft CA of $N$ ``ontic'' sites (``cogwheel mode'').  The relativistic coordinate perametrising this cyclic dynamics is named here ``ontic'' time $t$. The  EC infinitesimal evolution law in the ``ontic'' time is in general $| t \rangle \rightarrow | t + \delta t + \text{mod} ~T \rangle$.  

The periodic dynamics of an EC is described, at a statistical level, by a wave-function  $\Phi(t)$, also named ``physical state'', satisfying the Periodic Boundary Conditions (PBCs) $\Phi(t) = \Phi(t + T)$.  An EC can be effectively regarded as of a continuous periodic CA in which CA evolution law is encoded in PBCs. The essential idea is that ``\emph{there is a close relationship between a particle moving on a circle with period $T$ and the Quantum Harmonic Oscillator (QHO) with the same period}" \cite{'tHooft:2001ar}. This is a fundamental correspondence as the QHO is the essential ingredient of a second quantised field, and in turn of the whole Quantum Field Theory (QFT). In this paper we will use this correspondence to derive applications to condensed matter. 

 The powerful formalism of ECs theory allows for a direct test, in condensed matter, of advanced aspects of theoretical physics  \cite{Dolce:SuperC,Dolce:EPJP,Dolce:cycles,Dolce:AdSCFT,Dolce:tune,Dolce:EC2CA}. To every EC of period $T$ is associated a fundamental energy $\hbar \omega = 2 \pi \hbar / T $, according to the de Broglie phase harmony condition in the ``ontic'' time $\omega T = 2 \pi$. EC is characterised by undulatory mechanics so that the generic solution constituting the EC physical state has the form $\phi(t) = e^{-i \omega t}$. The PBCs $\Phi(t) = \Phi(t + T)$  encoding the EC evolution law in time determine the EC eigenstates $\phi_n(t)= e^{-i \omega_n t}/\sqrt{2 \pi}$.  They form a complete, orthogonal set with  harmonic energy spectrum $\hbar \omega_n = n \hbar \omega = n \frac{2 \pi \hbar}{T}$ ($n \in \mathbb Z$). Hence a classical EC naturally defines an Hilbert space of basis $| n \rangle$ such that $\langle t | n \rangle = \phi_n(t)$  with induced inner product $\langle n | n' \rangle = \delta_{n,n'}$. Thus an EC turns out to be represented  by the superposition of its eigenstates $\Phi(t) = \sum_{n \in \mathbb Z} \alpha_n \phi_n(t)$. That is, in the Hilbert space representation, it is described by a point in the corresponding Hilbert space $|\Phi \rangle = \sum_n \alpha_n |n \rangle$.  Since $i  \partial_t \phi_n(t) = \omega_n \phi_n(t)$ the EC ``ontic'' time evolution is given,  in this Hilbert space formalism, by the ordinary Schr\"odinger equation $i \hbar \partial_t  |\Phi(t) \rangle = \mathcal H  |\Phi(t) \rangle$ where the Hermitian operator $\mathcal H$ is defined as $\mathcal H | n \rangle = \hbar \omega_n | n \rangle$. Hence, its evolution is given by the Hilbert operator $\mathcal U(dt) = e^{-i \mathcal H dt / \hbar}$.  We must bear in mind that the fundamental topology of an EC is that of the circle $\mathbb S^1$. This determines, by means of the PBCs, the quantum number $n$. 



Actually, as can be checked in the ECs theory, the temporal dynamics of an EC correspond to those of the time evolution of a Quantum Harmonic Oscillator (QHO) of period $T$, except for the fact that the quantum number can assume negative values, $n \in \mathbb Z$ (positive and negative frequencies). This seems to imply a non positively defined Hamiltonian operator for EC. Nevertheless, as also argued by 't Hooft, the negative modes can be consistently interpreted as the anti-particles of relativistic QM. In confirmation of this we will in fact find that in graphene physics these negative modes correspond to the holes in the Fermi sea, i.e. to the negative modes of the Carbon Nanotubes (CN) energy bands. Indeed, the Elementary Charge Carriers (ECCs) in a CN actually behave as massless ``particles on a circle'', i.e. they simulate ECs in the essence. In particular, ECCs in a CN with $N$ carbon atoms along the circumference corresponds to ECs of $N$ lattice sites  on a circle. 
We will use the covariant generalisation of this correspondence to derive in a simple way, and in agreement with present literature \cite{RevModPhys.79.677,deWoul:2012ed}, the essential electronic properties of CNs. CNs can be therefore used to test ECs theory (and in turn CA models) and thus foundational aspects of QM. 

Similarly we will show here that the EC evolution law, i.e. the PBCs of the physical state, yields the most foundational aspects of SuperConductivity (SC) such as: quantisation of the magnetic flux in unit of $\varphi_0/2 = h c / 2 e $, the Meissner effect, the Josephson effect, the Little-Park effect, etc. Indeed, the EC evolution directly implies discretised variations of the Goldstone field which in turn describe the explicit breaking of the ElectroMagnetic (EM) gauge invariance typical of SC \cite{Dolce:SuperC,Weinberg:1996kr}.  In short, the SC phenomenology will be derived by applying the simple requirement that the wave function of electrons in the superconductor must have closed space-time orbits, in analogy with the closed space-time orbits of electrons in the atomic orbitals (see Bohr atom or Bohr-Sommerfeld quantisation). According to this description, the electron in the atomic orbitals can be therefore regarded as being locally in a superconducting regime.  

\section{Testing Elementary Cycles theory by means of Carbon Nanotubes: Compton clock vibrational modes and controvariant modulations.}

As well-known, electrons in a graphene monolayer effectively behave as massless charged particles (with pseudo-spin), i.e. as relativistic particles traveling at the Fermi velocity $v_F$ (the analogous of the speed of light for graphene physics). Let us assume that the graphene layer is curled-up in the same direction of the electron motion. The result will be a massless ``particle moving on a circle'' at ``light'' speed $v_F$. Since the CN have a finite number $N$ of carbon atoms along the circumference  $C_h$, such a ``circle'' is on a lattice of $N$ sites. In short, the electron moving at velocity $v_F$ along a CN circumference  characterised by $N$ carbon atoms can be regarded as an EC with $N$ lattice sites on a circle and time period $T_C = C_h / v_F $. According to ECs theory this must be identified with the Compton period of the ECCs implying an effective mass for the electrons. Indeed CNs are real physical systems simulating ECs (and CA for some aspects).  For the sake of simplicity we start by investigating the continuous lattice limit. The effective CN electronic properties will be then obtained by passing to the lattice case. For instance the EC harmonic energy spectrum $\hbar \omega_n = n \hbar \omega =  \frac{ 2 \pi \hbar n}{ T}$ in the lattice case is $\hbar \omega_n  = \frac{N \hbar }{2 T } \sin (\frac{\pi n}{N})$. 

As a consequence of the compactification of one of the two spatial dimensions of the graphene layer, the resulting CN effective dynamics are effectively characterised by a single spatial dimension, i.e. the axial direction $x_\parallel$. In such an effective CN one-dimensional space the electron moving along the circumference is effectively at rest with respect to the axial dimension. We denote the momentum of the electron along the axial direct with $\hbar k_\parallel$. The period of the electron moving along the CN circumference, which is effectively at rest with respect to the axial direction, is denoted by  $T(k_\parallel = 0) = T_C$. According to ECs theory, this effective rest period corresponds to a rest energy $\hbar \omega(k_\parallel = 0)$, such that  $T(k_\parallel = 0) = 2 \pi / \omega(k_\parallel = 0)$. In relativity it is natural to associate a rest energy to a mass: $\hbar \omega(k_\parallel = 0) = m_* v_F^2$. Indeed the rest period $T_C$ corresponds to the effective Compton time of the electron in the CN, which in turn fixes the effective mass scale of the ECCs in CN according to the Compton relation $m_* = \frac{2 \pi \hbar} {T_C v_F^2}$ (that is, the Compton time of a relativistic particle of mass $m$ is $T_C = \frac{2 \pi \hbar}{ m c^2}$). A massless particle generates an effective mass if constrained in a (periodic) box! \cite{'thooft:2015}. Thus, we find that the effective mass scale of the electrons in the CNs is determined by the CN circumference according to 
\begin{equation}
m_* = \frac{2 \pi \hbar}{ C_h v_F}
\end{equation}
This effective mass scale is the fundamental rest energy of our ECs of rest period $T(k_\parallel = 0) = T_C$. The ECCs at rest along the CN axial direction can be regarded as ECs of rest period $T_C$. They simulate the effective de Broglie internal clocks of the electrons in the CNs. Notice that the Compton time of an ordinary electron $\sim 10^{-21} $s  is effectively rescaled to about $10^{-15} $s in CNs. Since the latter time scale is accessible to modern timekeepers, CNs allow for a indirect test of foundational quantum aspects of elementary particles, as pointed out for instance in \cite{Dolce:EPJP}.  

 Furthermore, due to the periodic dynamics, an EC (i.e. a relativistic continuous CA) is characterised not only by the fundamental energy considered so far, but also by a whole spectrum of energy levels. In the rest case the EC rest energy spectrum is $\hbar \omega_n(k_\parallel = 0) = n \hbar \omega(k_\parallel = 0) = n m_*$. Since this is a rest energy spectrum, it is natural to associate it to a mass spectrum $ m_n = n m_* = \frac{2 \pi n \hbar}{ C_h v_F}$, with $n \in \mathbb Z$.  This yields interesting analogies to the Kaluza-Klein theory as described in \cite{Dolce:AdSCFT}. 
Similarly to the general case, this harmonic spectrum denotes the eigenmodes associated to the PBCs in the proper time $\tau$ (i.e. the time coordinate for the electron at rest with respect to the axial direction) for the EC describing the ECC in the CN: For the rest EC we have $\Phi(\tau) = \Phi(\tau + T_C)$. In short these PBCs encode the evolution law of the EC evaluated at rest. In the derivation of the lattice limit we have to consider that, as a consequence of the hexagonal geometry of the graphene layer, two kind of PBCs are possible for the electron moving along the CN circumference. Besides the trivial PBCs described above, it is also possible to have configurations symmetric under lattice rotations of $\pm \frac{2}{3} \pi$. In this case the PBCs are thus twisted by a corresponding factor. That is, in general, we can write the PBCs of the ECCs in CNs as 
\begin{equation}
\Phi(\tau) = e^{i 2 \pi \alpha} \Phi (\tau + T_C)\,
\end{equation}
where $\alpha = 0, \pm \frac{1}{3}$. The evolution law of the ECs simulating the ECCs in CNs from an initial proper time $\tau_i$ to a final one $\tau_f$  is $| \tau_i \rangle \rightarrow e^{i 2 \pi \alpha} | \tau_f + \text{mod} ~T_C \rangle $. 
By considering this twist of the PBCs the resulting mass spectrum of the rest EC turns out to have a shift $\alpha m_*$: 
\begin{equation}
m_n = (n + \alpha) \frac{2 \pi \hbar}{T_C} = (n + \alpha)  m_*= (n + \alpha) \frac{2 \pi \hbar}{C_h v_F}~, ~~ (n \in \mathbb Z)\,.
\end{equation} 
This implies that for $\alpha = 0$ the CN is characterised by a massless mode $m_0 = m_* = 0$, whereas all the other modes are massive: $m_n = n  m_*$ with $n \neq 0$. Indeed this case describes metallic CNs. In the  case $\alpha = \pm \frac{1}{3}$ all the modes are massive. This in fact corresponds to semiconducting CNs whose characteristic fundamental mode has effective mass 
\begin{equation}
m_* =   \frac{2 \pi \hbar}{3 C_h v_F}\,.
\end{equation} 
in agreement with present literature \cite{RevModPhys.79.677}. 
The lattice limit correctly leads to the effective CN mass spectrum \cite{deWoul:2012ed}. We find that, for ZigZag (ZZ) and ArmChair (AC) CNs of $N$ carbon atoms in the perimeter, the resulting mass spectra turns out to be, respectively,
\begin{equation}
\text{ZZ)} ~~~ m_n = m_* \frac{N}{\pi} \sin\left (\frac{\pi n}{N} \right )~;~~~~~~~~~~~~~~~~~~~~~~~~~~ 
\text{AC)} ~~~ m_n = m_* \frac{N}{3 \pi}\left[ 1+  2 \cos(\frac{\pi n}{N})\right] ~.
\end{equation}   

In order to derive the CN energy bands we must consider the relativistic transformation of our EC of rest period $T_C$. This can be easily done in analogy with undulatory mechanics. Indeed an EC can be regarded as a ``de Broglie periodic phenomenon'', which at rest reproduces the so called ``de Broglie internal clock''.  As originally pointed out by de Broglie, the period and wavelength of a moving particle, and thus its energy and momentum, are determined, through  Lorentz transformations, by the rest periodicity, i.e. by the Compton period of a corresponding ``periodic phenomenon''. ECs theory shows that de Broglie idea of internal clocks of particles, at the base of modern undulatory mechanics and relativistic QM, directly applies to CNs.  

The case of ECC with a non zero velocity $v_\parallel$ along the CN axial direction  corresponds to an effective Lorentz boost in the effective CN space-time. That is, the effective period $T(k_\parallel)$ and wave-length $\lambda_\parallel$  of an electron with non-vanishing momentum  $\hbar k_\parallel$ in the axial direction of a CN  is obtained from the Compton time (rest periodicity) $T_C$ by means of the Lorentz transformation $T_C = \gamma_\parallel T(k_\parallel) - \beta_\parallel \gamma_\parallel \lambda_\parallel$ in the effective CN space-time, where the Lorentz factor is $\gamma_\parallel = 1 / \sqrt{1 - \beta_\parallel}$ with  $\beta_\parallel = v_\parallel/v_F$. In short, the motion along the CN is described by boosting the rest ECs described above. The effective Compton period $T_C= \frac{C_h}{ v_F}$  implies, by means of Lorentz transformations, a time period $T(k_\parallel)$ and a spatial recurrence $\lambda_\parallel$  when the electron moves along the axial direction --- as a consequence, for instance, of an electric potential. These resulting recurrences in time and space determine the effective energy $\hbar \omega(k_\parallel) =  \frac{2 \pi \hbar}{T(k_\parallel)}  $ and the momentum $\hbar  k_\parallel = \frac{2 \pi \hbar}{\lambda_\parallel}$ of the electron in the CN, according to undulatory mechanics. Notice that the former relation is the EC phase harmony relation in time for a moving ECC in the CN as it relates the time period and  fundamental energy for an EC of momentum $k_\parallel$. In analogy to relativistic notations, by introducing the two tangent two-vector $k_\mu = \{\omega/v_F, - k_\parallel\}$ (covariant) and $\lambda^\mu = \{T v_F, \lambda_\parallel\}$ (controvariant),  the phase harmony relation can be generalised to the invariant form $m_* v_F^2 T_C / \hbar =  k_\mu \lambda^\mu = 2 \pi $. Thus, the relativistic constraints for these two vectors are respectively $m_* v_F^2 = \hbar^2 k_\mu k^\mu$ and $\frac{1}{T_C^2} = v_F^2 \frac{1}{\lambda^\mu} \frac{1}{\lambda_\mu}$, denoting the dual relativistic dispersion relations for the fundamental energy and time period, respectively. The covariant EC evolution law describing an electron moving in the effective CN space-time $x^\mu = \{t, x_\parallel \}$ is thus given by  $| x^\mu_i \rangle \rightarrow e^{i 2 \pi \alpha} | x^\mu_f + \text{mod} ~ \lambda^\mu \rangle $ (the EC evolution law in time and space is given by the component $\mu = 0$ and $\mu=1$, respectively).
In analogy to the relativistic Doppler effect, this implies that the time period $T(k_\parallel) = 2 \pi / \omega(k_\parallel) $ varies with $k_\parallel$ in such a way  that the electron fundamental energy turns out to satisfy the effective relativistic dispersion relation 
\begin{equation}
\hbar^2 \omega^2(k_\parallel) = \frac{(2 \pi \hbar)^2}{T^2(k_\parallel)} = m^2_* v_F^4 + \hbar^2 k_\parallel^2 v_F^2\,.
\end{equation} 
We have obtained, from simple considerations about the periodicity, that the motion of the electrons along CNs is the analogous of the relativistic motion, provided that the mass of the electron is replaced by the  effective mass $m_*$ and the speed of light by the Fermi velocity $v_F$.

By applying this effective relativistic dispersion relation to every mass eigenmode $m_n$ allowed by the electron effective Compton periodicity $T_C$ we immediately obtain the energy bands structure of CNs in the continuous limit
\begin{equation}
\hbar \omega_n(k_\parallel) = (n + \alpha)\frac{2 \pi \hbar}{T(k_\parallel)} = (n + \alpha) \sqrt{m^2_* v_F^4 + \hbar^2 k_\parallel^2 v_F^2}~,~~(n \in \mathbb Z)
\end{equation} 
The CNs energy bands follow from this result by performing the lattice limit (and considering that the CN lattice has also a periodicity along the axial direction characterising  Brillouin zones). Thus we have to consider an EC on a lattice (or the covariant generalisation of a periodic CA). For instance,  in agreement we present literature, in the case ZZ CNs we find the energy bands structure --- or similarly the density of states \cite{Dolce:EPJP}: 
\begin{equation}
\text{ZZ)} ~~~ \hbar^2 \omega_n(k_\parallel) = m_*^2 v_F^4 \frac{N^2}{3 \pi^2} \left( 1 + 2 \sin \frac{\pi n}{N}\right)^2 - \frac{2}{3} \hbar^2 k_\parallel^2 v_F^2  \cos \frac{\pi n}{N}\,.
\end{equation}  
 
This description is very interesting to test the validity of ECs theory (and CA models). Indeed one of the main issues of the CA interpretation of QM is that the negative modes ($n= -1, -2, \dots$) associated to the CA evolution law implies a non positively defined Hamiltonian operator. But ECs must be interpreted as relativistic objects (covariant generalisation of CA).  As also noticed by 't Hooft, these negative modes can be consistently interpreted as antiparticles, i.e. holes in the Dirac sea. This is fully confirmed by CNs physics in which, actually, the negative energy bands corresponding to the negative ECs vibrational modes, actually describes holes in the Fermi sea. Thus, in ECs theory the negative modes describe antiparticles.  

\section{Testing Elementary Cycles theory by means of superconductivity: the role of the temperature}

We now introduce the role of the temperature in ECs theory; it can also be generalised to CA. We must remember that in ECs physics, as in undulatory mechanics, the time period and the energy are ``two faces of the same coin''. Therefore interactions, i.e. local variations of energy $\hbar \omega = 2 \pi \hbar / T$, correspond to local modulations of the period $T$. Since  the temperature implies random collisions among particles (thermal noise), a quantum system at finite temperature is characterised by chaotic (Poissonian) breaks of the particle ``complete coherences''  with characteristic thermal time $\beta = \hbar / k_B \mathcal T$, i.e. a dumping $e^{-\hbar \omega / k_B \mathcal T} = e^{- 2 \pi \beta / T}$ of the cyclic behaviours of  periods $T$, being $k_B$ the Boltzmann constant and $\mathcal T$ the temperature. 

While a free EC  at zero temperature, as well as the perfect coherence of pure quantum systems, are characterised by perfect  periodicity $T$ in the Minkowskian time, thermal quantum systems are characterised by Euclidean time periodicity of period $\beta = \hbar / k_B \mathcal T$. These two kinds of temporal periodicities (which can be related by Wick's rotation, as for instance in the Hawking theory of black holes) have fundamental different physical meanings: they are in competition. In the pure quantum systems at zero temperature we have a persistent exact recurrence in time, e.g. described by the phasor $e^{-i \omega_n t}$, whereas in the latter case, corresponding to a non zero temperature, we have a dissipation of the quantum recurrence by Boltzmann factors $e^{-\omega_n / k_B \mathcal T} = e^{-2 \pi n \beta / T}$. That is  a dumping factor of the Minkowskian periodicity. In other words, while the Minkowskian periodicity $T$ of QM tends to form perfect coherent states (``periodic phenomena''), the Euclidean periodicity $\beta$ describes a dumping associated to the thermal noise which tends to destroy the perfect recurrences of  pure quantum systems. Thus, if the $T \ll \beta$ the system can autocorrelated and give rise to pure quantum phenomena such as SC. In the opposite limit the thermal noise breaks the the quantum recurrence before it can give autocorrelation, in this case we have the classical limit, e.g. ordinary electric resistance. 

Another aspect of the temperature which will be used in the following description is the Boltzmann distribution. In a few words, the dumping factor $e^{-\omega_n / k_B \mathcal T} = e^{-2 \pi n \beta / T}$ can be also regarded as the probability to populate the n$^{th}$ vibrational mode of the system. Thus, at very low temperature $T \ll \beta$ only the fundamental mode $n=1$ will be populated whereas at high temperature many vibrational modes must be considered (e.g. as in the classical limit in Bohr atom). We will use this aspect to describe the EM symmetry breaking occurring in superconductivity (in analogy with the effective gauge symmetry breaking  in extra dimensional extensions of the Standard Model, and bosonic condensation). 

These simple arguments, further motivated below, allow us to predict that high temperature SC happens in quantum systems characterised by very small quantum recurrences $T$, orders of magnitude smaller than in ordinary SC. This is confirmed by CNs in which, as we have seen, the quantum recurrence is explicitly determined by the CN circumference $C_h$, so that, in agreement with \cite{RevModPhys.79.677,White:nature}, the critical temperature is inversely proportional to the circumference $C_h$.

Our description of CNs shows that covariant ECs provide a model describing the ``periodic phenomena'' conjectured by de Broglie to describe the quantum behaviour of elementary particles. Indeed an EC of rest period $T_C$ describes the quantum behaviour of a relativistic particle of mass $m = 2 \pi \hbar / T_C c^2$, where we have replaced the Fermi velocity with the speed of light $c$. In general an EC, similarly to the ideal case of a free quantum particle at zero temperature, is characterised by a ``complete coherence'' in space-time. 

Generalising the CN description to the four-dimensional space-time of ordinary relativity, in a given inertial reference frame the free EC has time period $T(\vec k)$ and wave-length $\vec \lambda$ which determines its energy $\hbar \omega(\vec k) = 2 \pi \hbar / T(\vec k)$ and momentum $\hbar k_i = 2 \pi \hbar / \lambda^i$ ($i = 1,2,3$). In covariant notation we can introduce the four vectors $k_\mu = \{ \omega  / c , - \vec k\}$ and $\lambda^\mu = \{T c , \vec \lambda \}$ related by the phase harmony relation $k_\mu \lambda^\mu = 2 \pi \hbar$.  This means that the evolution law is $| x^\mu_i \rangle \rightarrow  | x^\mu_f + \text{mod} ~ \lambda^\mu \rangle $ (notice that in this case a twist fact of $\alpha = 1 / 2$ would reproduce the vacuum energy  $\hbar \omega(\vec k) / 2$). The generic EC solution, in the free case, is $\Phi(x) = \sum_n  e^{-i {k_{ \mu}}_n x^\mu}$ and the PBCs, encoding the evolution law, are $\Phi(x^\mu) = \Phi(x^\mu + \lambda^\mu)$. These imply the (normally ordered) energy-momentum spectrum $\vec {k_{\mu}}_n = n  \vec k_\mu$ of a free particle. As can be easily seen from EC time  evolution, $\Phi(t) = \sum_n e^{i \omega_n t}$ and $\Phi(t) = \Phi(t + T)$ imply the ordinary (normally ordered) energy spectrum of a relativistic quantum particle $\omega(\vec k ) = n  \hbar \omega(\vec k )$ (in analogy with a string vibrating with period $T(\vec k)$). In short, the EC evolution law implies that the physical state $\Phi$ describing the wave function of a free relativistic particle at zero temperature must have closed orbits along its space-time evolution. That is, we have a correspondence with the Bohr-Sommerfeld quantisation. Indeed, an interacting EC is characterised by a locally modulated space-time period so that, in the interacting case, the generic solution has the form of a locally modulated wave $\phi(x) = e^{- i \int^x d y^\mu k^\mu(y)}$. The evolution law, i.e. the local PBCs for this generic solution, in the interacting case yields the relativistic Bohr-Sommerfeld quantisation condition $\oint_{\lambda} d y^\mu {k_\mu}_n(y) = 2 \pi n \hbar$. It is well-known that in a Coulomb potential this yields the Bohr energy levels of the atomic orbitals $E_n = -13. 6 ~\text{eV} / n^2$.  

Let us assume unitary gauge invariance for an EC (representing electrons in equilibrium with the EM field in the superconducting material)
$ 
\Phi(\vec x, t) = U(\vec x, t) \Phi(\vec x, t)$, where $U(\vec x, t) = e^{- i \frac{e}{\hbar c} \theta(\vec x, t)}
$. 
The Goldstone $\theta(\vec x, t)$ denotes the local invariance of the EM interaction and $e$ is the electric charge.  In analogy with the EC description, we now impose PBCs  $\Phi(\vec x, t) = \Phi(\vec x, t + T(\vec k)) $ (for the sake of simplicity we only consider the time component, furthermore we assume here that the topology is the orbifold $\mathbb S^1/ \mathcal Z_2$ associated to the fundamental topology $\mathbb S^1$ of the EC, so that the phase invariance is $n \pi$ rather than $2 \pi n$). Thus, in this case the PBCs in time of period $T(\vec k)$ imply the following evolution law for the Goldstone mode
 \begin{equation}
\frac{e}{\hbar c}\theta(\vec x, t) = \frac{e}{\hbar c}\theta(\vec x, t + T(\vec k)) + \text{mod}~~ n \pi 
. \end{equation} 
As a consequence of the EC evolution law the Goldstone field can only vary by finite amounts 
\begin{equation}
\Delta \theta(\vec x, t) = \frac{\varphi_0}{2}~,~~\text{with}~~ \varphi_0 = \frac{h c}{e}\,.\label{discr:Gold}
\end{equation}

This simple condition yields the characteristic quantum behaviours of a superconducting regime, as described in \cite{Weinberg:1996kr}. In short, according to our ansatz, as a consequence of the closed orbits of the wave-function (physical state) imposed by the EC periodic dynamics (i.e. by the CA evolution law),  ${\theta}(\mathbf{x},t)$ can only vary
by discrete amounts running along a conductor contour $\Sigma$ in which the EM field is a pure gauge $A_\mu (\vec x, t) = \partial_\mu \theta(\vec x, t)$. The Stokes theorem gives a quantisation
of the magnetic flux through the area $\mathcal{S}_{\Sigma}$ limited by $\Sigma$,  
\begin{eqnarray}
 \int_{\mathcal{S}_{\Sigma}}\!\! \mathbf{B}(\mathbf{x},t)\cdot d\mathbf{S} 
= \oint_{\Sigma_{\mathcal{}}}\!\! \mathbf{A}(\mathbf{x},t)\cdot d\mathbf{x} 
  =  \oint_{\Sigma}\!\! \mathbf{\nabla}{{\theta}}(\mathbf{x},t)\cdot d\mathbf{x}=n \frac{   \varphi_{0}}{2 } 
  \,. 
  \label{magn:quant}
  \end{eqnarray}
 Since the magnetic flux is quantised, the electric current cannot
smoothly decay while flowing around the circuit, and there is not electric resistance, i.e. we have SC,  
\cite{Weinberg:1996kr} and \cite{Dolce:cycles,Dolce:2009ce,Dolce:tune}. 
The quantisation (\ref{discr:Gold})   means that 
the Goldstone field $\theta(\mathbf x, t)$ transforms as the phase of a condensate of a fermionic pair operator $\left\langle\epsilon_{\alpha \beta} \Phi^{\alpha} \Phi^{\beta}\right\rangle$  of charge $-2e$.   That is, it plays the role of the usual Cooper pair of the BCS microscopic theory of SC. From (\ref{discr:Gold})  it is also manifest that the quantisation of the Goldstone field   implies a breaking of the EM gauge $U_{EM}(1)$  to $\mathcal Z_{2}$.

Now we consider the EM gauge connection for the EC physical state $\Phi(x' ) = e^{i \frac{e}{\hbar c} \int_x^{x'} A_\mu d x^\mu} \Phi(x' )$. Indeed, gauge interaction has a particular geometrodynamical meaning in  ECs theory \cite{Dolce:tune}. This can be alternatively obtained from the minimal substitution $\hbar k_\mu(x) \rightarrow \hbar k_\mu(x) - e A_\mu(x) $ in the EC interacting solution.   By imposing the EC evolution law  in space-time with orbifold topology $\mathbb S^1/\mathcal Z_2$, i.e. the EC PBCs in space-time, this yields the Dirac quantisation condition for magnetic monopoles (in analogy with the Bohr-Sommerfeld quantisation)
\begin{equation}
\oint A_\mu d x^\mu = n \frac{\varphi_0}{2}\,.
\end{equation} 
The spatial component yields the quantisation of the magnetic flux described above whereas the time component yields the Josephson effect. Let us consider a junction between superconductors with a voltage difference $\Delta V$ (the BCs in this case is determined by the isolating barrier). Since the Maxwell equation in this case implies that $\Delta V = - \partial_t \theta$, the Stokes' theorem yields $T_{junct} \Delta V / c = \varphi_0 / 2$ corresponding to the frequency $f_{junct} = 1 / T_{junct}$ of the  Josephson effect. Alternatively this can be derived by considering that the Lagrangian at the junction can only depend on the phase difference of the Goldstone fields $\mathcal L_{junct} = \mathcal F (\Delta \theta)$. The Josephson effect follows from the fact that the evolution law implies that the Goldstone can only vary by finite steps, see eq.(\ref{discr:Gold}), so that $ \mathcal F (\Delta \theta) = \mathcal F (\Delta \theta + n \varphi_0 /2)$ \cite{Weinberg:1996kr}. 

Due to the quantum recurrence, which in our ECs theory originates from the PBCs, i.e. from the EC evolution law,  the EM gauge field $A_\mu$, similarly to the Goldstone field, is constrained to have a periodic behaviour. This means that it can be expanded in vibrational eigenmodes which, for the sake of simplicity, we assume here to be harmonic $A_\mu(\vec x,t) = \sum_{n \in \mathbb Z} e^{-i n \omega^\gamma t} A_{(n)\mu}(\vec x)$. This implies that the EM Lagrangian can be expanded in vibrational eigenmodes as well. In the temporal (unitary) gauge we get
\begin{equation}
\mathcal L_{YM} = - \frac{1}{4} \int d^4 x   F_{\mu \nu}  F^{\mu \nu} =  \sum_{n \in \mathbb Z}  \int d^3 x \left[- \frac{1}{4}   F_{(n) \mu \nu}  F_{(n)}^{ \mu \nu} -  \frac{1}{2} \frac{(\omega_n^\gamma)^2}{c^2} A_{(n)\mu}A_{(n)}^{\mu} \right]
\end{equation}  
 However we must consider that at very low temperature $T \ll \beta$ only the lower mode of the system, that here we will denote by the bar sign, is populated. This means that the summation of the above Lagrangian must be truncated to the fundamental mode. As well-known for gauge invariance in extra dimensional theories (see Higgsless, Composite Higgs or unified gauge-Higgs models), this implies that the low temperature is no more gauge invariant. So we find again an effective EM gauge breaking. Roughly speaking, at very low temperature we can imagine that, similarly to a Bose-Einstein condensation, all the vibrational modes of energy $\omega_n^\gamma = n \omega^\gamma$ condensate on the fundamental mode. 
 
 This symmetry breaking can also be seen from the effective matter Lagrangian $ \mathcal L_S$ obtained by integrating out from the equations of motion the contributions of the higher modes as a function of the fundamental one. By expanding $ \mathcal L_S$ for small perturbations around the fundamental mode, which now can be effectively written as $\bar \Phi = \rho e^{i 2 e \theta}$, we get  
 \begin{eqnarray}
  \mathcal L_S &\simeq& \int d^3 x \left [ - \frac 1 2 \bar \Phi^* |(\vec \nabla - i 2 e \vec A)|^2  \bar \Phi - \frac 1 2 \alpha  |\bar \Phi|^2   - \frac 1 4 \beta |\hat \Phi|^4 \right] \nonumber \\
  &=& \int d^3 x \left [ - 2 e^2 \rho^2 (\vec \nabla - i 2 e \vec A)^2 - \frac 1 2 (\vec \nabla \rho )^2 - \frac 1 2 \alpha \rho^2 - \frac 1 4 \beta \rho^4 \right ]\,. 
   \end{eqnarray}
   This describes (with appropriate normalisation) the BCS theory of ordinary SC. 
The first term describes the fundamental vibrational mode without the perturbation of the higher modes.  Due to gauge transformations it can be written as $\bar \mathcal L_S[A_\mu - \partial_\mu \phi] \sim \frac{V}{\Lambda^2} (A_\mu - \partial_\mu \theta)$, $V$ is the volume and  $\Lambda = \frac{1}{\sqrt{4 e^2 \langle \rho^2 \rangle}}$ is the penetration length  of the Meissner effect.   
Indeed, deep inside the superconductor where the EM field is pure gauge, we have  $F_{\mu \nu} (\partial_\chi \theta) = 0$ and the magnetic field is vanishing $\vec B = 0$. $\bar \mathcal L_S$ describes the fact that  the energy cost to expel the magnetic field is small with respect to  that of the classical configuration. By considering that the minimum of $\rho$ is $\langle \rho^2 \rangle = - \frac{\alpha}{\beta}$ we get $\Lambda = \frac{1}{2 e}\sqrt{- \frac{\alpha}{\beta}}$, in agreement with ordinary  SC. Finally, by expanding  $\rho$ around its minimum $\rho = \langle \rho \rangle +  \rho'$ we get $\vec \nabla^2 \rho' = - 2 \alpha \rho'$. It allows us to introduce the coherence length $\xi = \frac{1}{- 2 \alpha}$ describing the stability of the vortices and of the flux-quantised magnetic field. Indeed the energy density of the superconducting state is lower than the energy density of the normal state.  We have $\frac{\alpha^2}{4 \beta} = \frac{1}{32 e^2 \lambda^2 \xi^2}$.   Thus, in agreement with the ordinary description, the case $\xi > \Lambda$ and  $\xi < \Lambda$ describe Type I SC and Type II SC, respectively. 

\section{Conclusions}

  In this paper we have shown that  the interest of the Elementary Cycles theory,  or similarly 't Hooft Cellular Automata models, is not only limited to our knowledge of foundational aspects of Quantum Mechanics: it also represents a powerful tool to derive and describe in a very simple way nontrivial quantum phenomena of condensed matter, such as the behaviour of the Elementary Charge Carriers in Carbon Nanotubes and the essential phenomenology of Superconductivity.  Indeed we have derived the Carbon Nanotubes (or similar graphene systems) electronic properties by simply considering the relativistic modulations of the vibrational modes characterising  the cyclic dynamics of Elementary Cycles. Similarly we have derived all the fundamental quantum phenomenology associated to Superconductivity and the role of the temperature in quantum systems, particularly relevant to interpret the occurrence of High Temperature Superconductivity. These represent further confirmations of the overall validity of the Elementary Cycles formulation of Quantum Mechanics.

\ack
We acknowledge G. 't Hooft for useful comments and   partial financial
support from the project
FAR ``\emph{Control and enhancement of superconductivity by
engineering materials at the nanoscale}''.

\section*{References}


\providecommand{\newblock}{}

%
%
%
%
%
%
%
%
%
%
%
%
%
%

\end{document}